\PassOptionsToPackage{table}{xcolor}

\documentclass[sigconf,screen,nonacm]{acmart}

\usepackage{booktabs}
\usepackage{graphicx}
\usepackage{longtable}
\usepackage{array}
\usepackage{tabularx}
\usepackage{makecell}
\usepackage{multirow}
\usepackage{rotating}
\usepackage{tikz}
\usepackage[skins]{tcolorbox}
\usepackage{xcolor}

\definecolor{criteriaBlue}{HTML}{C9D9E8}
\definecolor{criteriaGreen}{HTML}{D4E2C7}
\definecolor{criteriaPurple}{HTML}{EED3BD}

\newcolumntype{L}{>{\raggedright\arraybackslash}X}

\newcommand{\paratitle}[1]{\vspace{1ex}\noindent\textbf{#1}}

\setcopyright{cc}
\setcctype{by}
\copyrightyear{2026}
\acmYear{2026}
\acmDOI{XXXXXXX.XXXXXXX}

\begin{document}

\title{Defining Page-Centric Visual Document Retrieval and Evaluating Existing Benchmarks}

\author{Chongzhi Zhang}
\email{chongzhi001@e.ntu.edu.sg}
\affiliation{%
  \institution{Nanyang Technological University}
  \city{Singapore}
  \country{Singapore}
}
\author{Yuqi Zuo}
\email{zuoy0004@e.ntu.edu.sg}
\affiliation{%
  \institution{Nanyang Technological University}
  \city{Singapore}
  \country{Singapore}
}
\author{Aixin Sun}
\email{axsun@ntu.edu.sg}
\affiliation{%
  \institution{Nanyang Technological University}
  \city{Singapore}
  \country{Singapore}
}

\begin{abstract}
In modern information retrieval (IR), documents are increasingly encountered as structured sequences of visually rich, paginated units, rather than as flat text. Example documents encountered in our daily life include Word files, PDFs, and slide decks. Free text-centric retrieval systems either ignore this structure entirely or destroy it through arbitrary passage chunking. In this paper, we formulate page-centric Visual Document Retrieval (VDR) with pages as the fundamental indexing units and document context available for assessing page relevance. The formulation distinguishes two retrieval granularities, page and document, and three retrieval paradigms: independent-item, evidence-set, and iterative retrieval.  We then assess 42 benchmarks using eight criteria covering corpus characteristics, query construction, relevance annotation, and ranking support. Our analysis shows two main findings on current VDR research. First, evaluation remains concentrated on page-level independent-item retrieval, with fewer resources supporting document-level, evidence-set, and iterative settings. Second, queries originating from real user information needs independently of target corpus content and graded relevance judgments are uncommon, while the coverage and verification of relevance judgments vary substantially. Both findings reveal gaps between the capabilities applications require and those existing evaluation protocols can establish. We provide benchmark selection guidance  and identify priorities for future benchmark design, emphasizing alignment between task requirements, relevance semantics, and evaluation protocols.
\end{abstract}

\ccsdesc[500]{Information systems~Test collections}
\ccsdesc[500]{Information systems~Relevance assessment}
\ccsdesc[500]{Information systems~Document filtering}

\keywords{Visual document retrieval, benchmark analysis, multi-modal retrieval, retrieval-augmented generation}

\maketitle

\section{Introduction}

Real-world document corpora such as technical reports, slide decks, scanned forms, financial filings, are rarely flat text. They are organized as ordered sequences of visually formatted pages. The layout, typography, and embedded figures or tables carry meaning that is inseparable from the surrounding text in each or a few pages in sequence. Traditional text-centric IR systems handle such corpora poorly: they either treat a document as an unstructured stream of text or rely on arbitrary passage chunking, both of which discard the spatial pagination and visual structure that document authors use to convey relevance-bearing information~\cite{tito2023mpdocvqa}. 

In fact, this points to a more basic mismatch. In classical IR, a "document" is a \textit{logical unit} of information defined for the purposes of indexing and retrieval: a convenient abstraction, not a claim about how information is actually packaged. That abstraction often diverges from the documents we produce and read day to day, which are \textit{physically paginated},\footnote{Webpages are considered a special case because they can be treated as single-page documents when viewed online or as paginated documents when printed.} \textit{visually structured} artifacts. This calls for serious consideration of a page-centric view of information retrieval: treating the page rather than the document as the fundamental unit of indexing and retrieval. Accordingly, we examine the applications enabled by this view and ask how well page- and document-level retrieval serve end users’ information-seeking needs, particularly when documents span multiple pages and users need to refer to individual pages or entire documents.

Recent work on Visual Document Retrieval (VDR) has begun to address this gap by treating a \textit{page image} itself as the unit of retrieval, most notably through late-interaction multi-vector models evaluated on the ViDoRe benchmark family~\cite{ViDoRe-v1,ViDoRe-v2,ViDoRe-v3}. However, the literature has grown organically around whatever a given benchmark happens to measure, and no work to date separates the \emph{formal task} of page-centric visual document retrieval from the \emph{benchmark design choices} used to evaluate it. This conflation makes it difficult to answer two questions that matter for both research and deployment: (1) what precisely is being retrieved and ranked, for example, individual pages, whole documents, or both, and (2) which of the many existing benchmarks are actually fit to evaluate which kinds of application. This paper makes three contributions:

\paratitle{A page-centric task formulation (\S\ref{sec:task}).} We explicitly define page- and document-level relevance,  and organize retrieval settings by granularity and retrieval paradigm, with motivating applications detailed in Appendix~\ref{sec:app-bench}.

\paratitle{A structured analysis of benchmark readiness (\S\ref{sec:benchmarkReadiness}--\S\ref{sec:scoring}).} We define 8 criteria and apply them to 42 benchmarks, clarifying which retrieval capabilities their corpus construction, annotations, and evaluation protocols can meaningfully assess.

\paratitle{Task-specific benchmark guidance (\S\ref{sec:scoring}, and detailed in Appendix~\ref{sec:app-bench}).} We structure our benchmark recommendations across two levels of granularity (i.e., page- and document-level relevance) and three retrieval paradigms (individual items, evidence set, and iterative retrieval), yielding six distinct granularity-paradigm combinations. For each combination, we outline application-based selection rules and discuss evaluation gaps and design considerations for future benchmark design.

\section{Related Work}
\label{sec:related}

In IR, documents are often represented and retrieved as whole documents or smaller passages using sparse or dense representations. Passage-level retrieval commonly relies on chunking documents into fixed-length token spans, disregarding their original layout. In VDR, pages serve a similar functional role as passages, providing sub-document units suitable for embedding, retrieval, and downstream processing, such as answer generation.

However, the two approaches differ in how retrieval boundaries are determined. Token-based chunking imposes artificial boundaries, often dictated by model constraints, which may split semantically coherent elements such as tables or figure captions. In contrast, page boundaries are inherent to fixed-layout documents, such as PDFs and slide decks, and often reflect layout decisions intended to preserve the semantic coherence of visual and textual content. Page-level retrieval thus leverages the document's existing layout structure rather than imposing arbitrary token-based boundaries. This makes pages a natural retrieval unit for visual documents, as reflected in the predominance of page-level evaluation among existing VDR benchmarks (\S\ref{ssec:cardinality}).

Nevertheless, page boundaries do not always strictly align with semantic boundaries. Identifying the relevance of an individual page may require contextual information from neighboring pages or the entire document. Document-aware passage retrieval provides a useful precedent for addressing such dependencies. DAPR~\cite{DAPR} and ConTEB~\cite{ConTEB} investigate text retrieval settings where identifying relevant passages requires information beyond the passages themselves. More recently, CMDR~\cite{CMDR} extends this perspective to visual documents by introducing context-dependent page retrieval and contextual multimodal embeddings within an intra-document retrieval setting. Building on these studies, our task formulation explicitly accounts for document context when assessing page relevance (\S\ref{sec:task}). Accordingly, our benchmark analysis examines whether existing evaluation settings can meaningfully assess such contextual dependencies (\S\ref{sec:scoring}).

\paratitle{Visual document retrieval benchmarks.} Early VDR benchmarks often repurposed document visual question answering (VQA) datasets by treating questions as retrieval queries and their source images as relevant targets, as exemplified by ViDoRe V1~\cite{ViDoRe-v1} and VisRAG~\cite{VisRAG}. Subsequent benchmarks have expanded evaluation along several dimensions, including multi-page and multi-document collections~\cite{M3DocVQA,ViDoRe-v3,MADQA,ViMDoc}, more realistic query construction~\cite{ViDoRe-v2,ViDoRe-v3,TaR-ViR}, graded relevance judgments~\cite{MultiDocR,ViDoRe-v3,KoViDoRe}, and complementary evidence retrieval across reasoning hops~\cite{Double-Bench}. More recent efforts further explore whole-document relevance~\cite{DocViRe} and iterative evidence acquisition~\cite{VisDocAgentBench,M4DocBench,PaperScope}. Collectively, these developments have broadened VDR evaluation beyond conventional single-item retrieval toward more diverse retrieval objectives and settings. Building on these advances, we systematically examine benchmark design choices in \S\ref{sec:benchmarkReadiness} and assess their suitability for different retrieval requirements in \S\ref{sec:scoring}.

\paratitle{Surveys and benchmark analysis.} Recent surveys have extensively reviewed VDR and document RAG from different perspectives. Zhang~\cite{VDR_Survey_Zhang} categorizes MLLMs by their roles in VDR as captioners, multimodal embedders, and end-to-end representers, examining retrieval granularity and fidelity--efficiency trade-offs. Gao et al.~\cite{Doc_RAG_Survey_Gao} organize document RAG models by domain, modality, and granularity, while summarizing benchmark characteristics, including corpus scale, content types, and reported retrieval and generation performance. Most relevant to ours, is Yan et al.~\cite{VDR_Survey_Yan}. In this dedicated survey of VDR, authors analyz existing benchmarks in terms of corpus and query sizes, multilingual coverage, reasoning requirements, and evaluation metrics under a page-image ranking formulation. They further distinguish retrieval methodologies from their applications in RAG and agentic systems. 

Not focusing on methodological aspects of VDR, our work complements these surveys by systematically assessing benchmark suitability against an explicit VDR task formulation and its application requirements. Specifically, we distinguish page-level and document-level ranking, and three retrieval paradigms: independent-item, evidence-set, and iterative retrieval. Based on this formulation and the practical applications, we establish 8 operational criteria covering corpus structure and scale, query construction, relevance coverage, result cardinality, relevance grades, and ranking support. We then evaluate 42 benchmarks against these criteria and use the resulting assessments to provide task-specific benchmark recommendations. Rather than merely cataloging benchmark characteristics, our analysis aims to clarify which retrieval capabilities each benchmark can meaningfully evaluate, establish a common vocabulary for describing benchmark properties and identifying potential evaluation gaps and biases.

\section{Task Formulation}
\label{sec:task}

Let $\mathcal{D} = \{D_1, D_2, \dots, D_N\}$ be a corpus of $N$ structured documents. Each document $D_i$ is an ordered sequence of formatted pages, $D_i = \left( p_{i1}, p_{i2}, \dots, p_{iM_i} \right)$, where $M_i$ is the number of pages in $D_i$. Each page $p_{ij}$ is a multi-modal unit $p_{ij} = \left( t_{ij}, v_{ij} \right)$, where $t_{ij}$ denotes textual content like body text, headers, captions, and $v_{ij}$ denotes visual characteristics. Examples of the latter include layout, typography, charts, tables, embedded figures, and spatial formatting. 

Note that the separation of $t_{ij}$ and $v_{ij}$ in defining a page $p_{ij}$ is for a formal description of page presentation as well as the comprehensiveness of the problem definition. It does not constrain whether a page must contain both modalities, or how a page is modeled during retrieval, e.g., as a page image capturing everything on the page or as a collection of textual and visual components parsed according to the page layout. 

\subsection{Page Relevance}
\label{ssec:pageRelevance}
Let $q$ denote a user's information need, expressed as a textual, visual, or multi-modal query. For simplicity, we consider only textual queries throughout this paper, as query modality is not our primary focus. The core computational objective of VDR is to learn a joint multi-modal relevance function
\begin{equation}
  s_{ij} = f\!\left(q,\, p_{ij} \mid D_i\right),
  \label{eqn:relevance}
\end{equation}
estimating the relevance of page $p_{ij}$ to query $q$, \emph{conditioned on its parent document} $D_i$ as global context.

In Eq.~\ref{eqn:relevance}, conditioning on $D_i$ allows $f$ to capture cross-page dependencies, document-level discourse, and sequential semantic flow that a page-isolated scorer cannot represent. Nevertheless, the conditioning on $D_i$ can be relaxed, or even completely dropped, so that $s_{ij}$ is estimated purely from the $p_{ij}$ alone and $q$. This relaxed form is worth distinguishing explicitly because it is what most existing retrieval systems actually implement: treating each page as an independent unit rather than reasoning over document-level context. We return to this distinction in our discussions later.

\paratitle{Relevance semantics.}
Eq.~\ref{eqn:relevance} produces a score. The ground truth it is evaluated against is a relevance label $\mathrm{rel}(q, p_{ij}) \in \{0, 1, \dots, L\}$, with $L = 1$ for binary judgments. Benchmarks give these labels different meanings, and three of them matter for the analysis in \S\ref{sec:benchmarkReadiness}.

A page is \emph{sufficient} if it contains the information needed to satisfy $q$ on its own. It is \emph{partially relevant}  if it contains necessary but incomplete evidence, such as one hop of a multi-hop question or the continuation of a table. It is \emph{contextual} if it contains little direct evidence but is needed to identify or disambiguate the target page, for example the page that introduces the entity the question refers to. Treating all three as equally “relevant” may obscure what a retriever is actually capable of doing.

Once the relevance score $s_{ij}$ is estimated, it is not difficult to localize and highlight some supporting evidences within a page, for example, as bounding boxes around relevant regions. We treat such highlighting as a post-processing step and not part of the essential retrieval process. The specific form of such evidence and how to present them to end users are application-dependent and not the key focus of our discussion here.

\begin{figure}
    \centering
    \includegraphics[width=0.95\linewidth, trim={1.5cm 12cm 10cm 4.5cm}, clip]{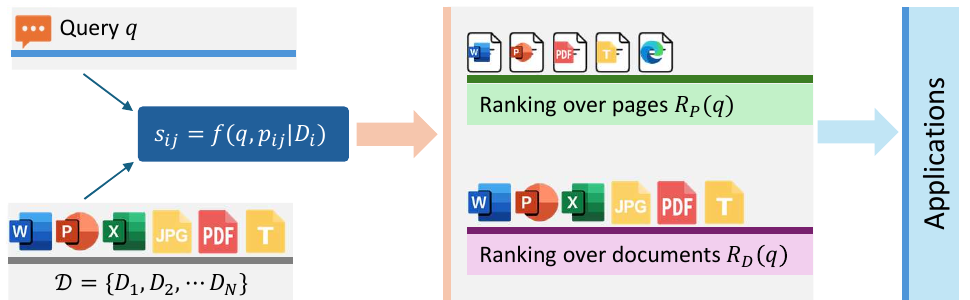}
    \caption{Overview of page-centric VDR. A document-conditioned relevance function $f$ scores each page of every document in the corpus; the scores support two rankings, over pages ($R_P(q)$) and over documents ($R_D(q)$), which serve the downstream applications.}
    \label{fig:scoring}
    \Description{A flowchart illustrating the page-centric visual document retrieval framework. A user query and a corpus of visual documents are fed into a document-conditioned relevance function, which assigns a relevance score to each page based on the query and its parent document. These scores support two retrieval outputs: a ranked list of individual pages and a ranked list of documents. Both rankings are subsequently used by downstream applications.}
\end{figure}

\subsection{Retrieval and Ranking Granularities}
\label{subsec:granularities}

Depending on the downstream application, VDR supports retrieval and ranking at two complementary granularities. Page-level ranking can surface relevant pages whose signal would be diluted or lost if only document-level scores were reported. Document-level ranking requires a non-trivial aggregation function that page-level ranking does not need.

\paratitle{Page-level ranking.}
The system returns a single global ranked list of individual pages across the entire collection,
\begin{equation}
  R_P(q) = \left( p^{(1)}, p^{(2)}, \dots \right), \quad
  \text{such that } s^{(1)} \geq s^{(2)} \geq \dots,
\end{equation}
where each retrieved page retains a deterministic pointer to its parent document, giving immediate access to broader context.

\paratitle{Document-level ranking.}
The system instead returns a ranked list of full documents,
\begin{equation}
  R_D(q) = \left( D^{(1)}, D^{(2)}, \dots \right),
\end{equation}
ordered by a document-level relevance score $S_i = g\!\left(\{s_{ij}\}_{j=1}^{M_i};\, D_i, \mathcal{D}\right)$ derived from an aggregation function $g(\cdot)$ (e.g., max-pooling, soft-top-$k$, or attention-based pooling) over constituent page scores. The arguments after the semicolon denote optional document attributes (e.g., length, title, metadata, authority measures) and corpus-level statistics. For each retrieved document, the system surfaces the specific attribution pages that drove its document-level rank.

\paratitle{Document relevance.}
The ground truth for document-level ranking can be defined in three ways. (a)~\emph{Existential}: $D_i$ is relevant if it contains at least one relevant page, $\mathrm{rel}(q, D_i) = \max_j \mathrm{rel}(q, p_{ij})$. (b)~\emph{Holistic}: $D_i$ is relevant because the document as a whole addresses the information need, e.g., the lease agreement or the deck that introduces a product. (c)~\emph{Evidence-set}: $D_i$ is relevant because it supplies one member of the set of documents needed to answer the query jointly. Under (a) and (c), document labels are deterministic functions of page labels, so document-level ranking reduces to page-level ranking followed by aggregation. Under (b), aggregation over page scores can fail in both directions: a document-level topical match may not show up in any single page score, while max-pooling over a long document can reward one incidental page.

\subsection{Retrieval Paradigms and Evaluation}
\label{ssec:paradigms}

Granularity specifies \emph{what} is ranked. A second, orthogonal axis specifies \emph{how} the ranked list is judged. Next, we distinguish three paradigms.

\paratitle{Independent items.} Each ranked page or document is evaluated independently based on its relevance label, even when several items are relevant to the same query. This mirrors the classic IR setting, and the standard IR metrics apply: nDCG@$K$ for graded relevance labels, and Recall@$K$, MAP, or MRR for binary relevance labels.

\paratitle{Evidence set.} Answering $q$ requires a set $E(q)$ of pages (or documents) that are jointly sufficient, for example one page per reasoning hop. The scorer $f$ may remain item-wise, but success is judged on the set: All@$K$ requires every member of $E(q)$ in the top $K$, whereas Hit@$K$ per hop requires at least one acceptable item for every hop and thereby allows substitutable evidence. We classify a benchmark under this paradigm only if it explicitly annotates complementary relationships among evidence items and adopts set-level evaluation metrics that assess their joint retrieval.

\paratitle{Iterative retrieval.} An agent interleaves planning, retrieval, and reasoning, so retrieval at step $t$ is conditioned on the history $h_t$ of earlier steps, $s^{(t)}_{ij} = f(q, p_{ij} \mid D_i, h_t)$. Evaluation targets the final evidence, and a shared protocol is still missing.

Each paradigm can be applied at page level or at document level, giving a $2 \times 3$ space of evaluation settings. For document-level ranking, a benchmark can also evaluate \emph{attribution}, i.e., whether the pages surfaced for a retrieved document include a relevant page. Existing benchmarks mostly populate a single cell of this space, i.e., independent-item evaluation at page level (\S\ref{ssec:cardinality}).

\section{Evaluating Existing VDR Benchmarks}
\label{sec:benchmarkReadiness}

A benchmark's ability to faithfully evaluate a task depends not merely on whether its input/output format matches the task's specification, but on whether the process by which it was constructed actually reflects the task's underlying requirements.  The comparison of SQuAD2.0~\cite{SQuAD2.0} and Natural Questions~\cite{Natural-Questions} in~\cite{DataReality23}illustrates this point: SQuAD2.0 contains questions written from passages that already contain the answer, whereas Natural Questions is based on real search queries posed before the answer is known. Consequently, despite sharing the same (question, passage, answer) format, the datasets represent different tasks, reflected in differences in lexical diversity, question specificity, and syntactic variety.

This distinction generalizes into two broadly different paths for benchmark construction.  The first \textit{repurposes an existing dataset} built for a different task, by reformatting it to match the input/output signature of the target task. This path is inexpensive, which is why many benchmarks accumulate around a convenient format rather than a carefully specified task, but it offers no guarantee that the repurposed dataset reflects the target task's actual constraints, query distribution, or notion of relevance.  The resulting benchmark may end up testing whether a model fits the idiosyncrasies of the source dataset rather than whether it solves the intended task. 

The second path \textit{works top-down from an explicit task definition}. The dataset is developed by enumerating the task's requirements first: who the intended users are, what constitutes a query, what relevance judgments should capture. Then, the collection and/or annotation process is guided to meet the task definition. This is the same discipline that motivates datasheets for datasets~\cite{gebru2020datasheets} and data statements~\cite{Bender2018datastatements}, which require a dataset's creators to document its motivation, the situation in which the underlying content was produced, and its collection and annotation process, precisely so that a downstream user can judge whether the dataset actually matches the task they intend to evaluate.

We evaluate 42 VDR-relevant benchmarks\footnote{We include only benchmarks that support the retrieval tasks defined in \S\ref{sec:task}, excluding those restricted to retrieval within one or a few predefined documents. We further require an explicitly defined retrieval task and corresponding evaluation metrics.} along three groups of dimensions: (i) document source and characteristics (\S\ref{ssec:sources}), (ii) query and relevance annotation processes (\S\ref{ssec:annotation}), and (iii) annotation results, covering both page- and document-level retrieval (\S\ref{ssec:cardinality}). Table~\ref{tab:scoringCriteria} details the 8 specific criteria under the three groups.

\subsection{Document Sources and Characteristics}
\label{ssec:sources}

Over time, the \emph{sources} and \emph{domain coverage} of VDR benchmarks have broadened. Early benchmarks often repurposed existing QA datasets or relied on relatively homogeneous collections~\cite{ViDoRe-v1,DocHaystack-InfoHaystack,VisRAG,OpenDocVQA}, whereas later benchmarks incorporate a wider range of professional materials from multiple disciplines~\cite{Double-Bench,UniDoc-Bench,FinMRAGBench,MADQA}. Multilingual coverage has also expanded beyond English to European~\cite{ViDoRe-v2,ViDoRe-v3,Double-Bench,Nayana-IR,JinaVDR}, East Asian~\cite{Double-Bench,KoViDoRe,SDS-KoPub-VDR,Nayana-IR,JinaVDR}, Southeast Asian~\cite{Nayana-IR,SeaDoc,JinaVDR}, and other languages~\cite{Double-Bench,Nayana-IR,JinaVDR}.

Document structure has received increasing attention, with evaluation extending from isolated page images or cropped elements to intact multi-page documents. Early benchmarks often treated figures, tables, or other extracted elements as standalone retrieval units~\cite{ViDoRe-v1,VisRAG,SciMMIR,TaR-ViR}, which is useful for evaluating understanding of specific content types but does not fully capture a model’s ability to integrate heterogeneous information within a document. There are also datasets consist of self-contained single-image documents, such as advertisements~\cite{CHaystack}, handwritten pages~\cite{BenthamQA-HW-SQuAD}, and invoices~\cite{Invoice-Haystack}. More recent benchmarks increasingly preserve complete multi-page documents, allowing retrieval to be evaluated in a more realistic structural context~\cite{ViDoRe-v3,MADQA,VisDocAgentBench}.

\paratitle{Corpus size and document length.} While paper-oriented collections average around 20 pages~\cite{IRPAPERS,VisDocAgentBench,Double-Bench}, long-document benchmarks can be substantially larger; for example, FinMRAGBench~\cite{FinMRAGBench} reports an average of 133.5 pages per document. \emph{Corpus size} varies even more widely, which can be measured either in number of pages or number of documents.  Candidate pools can reach tens or hundreds of thousands when pages or extracted elements are treated as retrieval units~\cite{SlideVQA-DSE,SciMMIR}. Others contain relatively few but much longer, carefully curated documents. Such collections can still create challenging retrieval settings because relevant evidence may span multiple pages or documents, as explicitly represented in KoViDoRe~\cite{KoViDoRe} and MADQA~\cite{MADQA}. For consistency, corpus size is measured by the number of pages in our scoring system.

\paratitle{Visual dependence.} Earlier datasets often assumed visual relevance indirectly through the input format or retrieval target, for example by retrieving figures, charts, or tables~\cite{ViDoRe-v1,VisRAG,SciMMIR}. More recent benchmarks increasingly annotate, control, or measure the modality of required evidence. ViDoRe V3~\cite{ViDoRe-v3} annotates evidence modality and provides region-level grounding in addition to page relevance. M4DocBench~\cite{M4DocBench} provides document-, page-, and region-level evidence annotations and identifies 112 of its 158 instances as multimodal. MADQA~\cite{MADQA} reports the proportion of visually dependent instances that requires understanding layout, tables, or other document artifacts. Overall, benchmark design has progressed from simply using document images, to retrieving explicitly visual content, and further toward directly characterizing the visual evidence required for successful retrieval.

However, we argue that benchmarks for generic visual document retrieval should not overemphasize non-textual visual evidence at the expense of textual content. In real-world documents, text remains a primary carrier of semantic information, and effective retrieval should leverage both textual and visual signals. Moreover, under our framework, the document page serves as the fundamental retrieval and indexing unit, with its layout, typography, and structural organization also contributing to the evidence available for retrieval. Therefore, we do not require every test case to involve non-textual visual evidence, nor do we score visual dependence. Instead, we recommend annotating evidence types (e.g., text, table, chart, and layout) to enable stratified evaluation.

\subsection{Source of Queries and Annotation Process}
\label{ssec:annotation}

We now focus on how query-item relevance is established and annotated. We first record the benchmark’s \emph{annotation direction}: \emph{query-to-source}, where existing user queries (e.g., from search logs) are matched to relevant corpus items; or \emph{source-to-query}, where queries are constructed by annotators or LLMs from selected corpus content to simulate user information needs. We further record the \emph{coverage of relevant items}, indicating whether the benchmark attempts to identify all relevant pages and documents for each query across the corpus.

\paratitle{Query-to-source vs Source-to-query.} Among all evaluated benchmarks, only TaR-ViR~\cite{TaR-ViR} adopts a query-to-source \emph{annotation direction}. Specifically, it reuses queries from the textual retrieval dataset NQ-TABLES~\cite{NQ-Tables}, whose queries originate from the Google search logs used in Natural Questions~\cite{Natural-Questions}, and identifies relevant items for those queries within a newly constructed multimodal corpus.

All remaining benchmarks follow a source-to-query paradigm, in which human annotators or LLMs generate queries after inspecting the corpus content. This process introduces two potential biases. First, the resulting queries may not reflect how users express information needs in practice, where queries are often underspecified, informal, and shaped by varying levels of domain knowledge~\cite{kdd_paper}. As a result, benchmark performance may not accurately predict real-world retrieval effectiveness. Second, because source-to-query construction begins with a known source, it guarantees the existence of supporting evidence by design. This may underrepresent realistic scenarios in which the corpus contains only partially relevant evidence or no relevant item at all.

Although source-to-query annotation is widely used because real search queries are difficult to obtain, some benchmarks explicitly seek to reduce its associated bias. For example, ViDoRe V2~\cite{ViDoRe-v2} and V3~\cite{ViDoRe-v3} recognize that annotators often have access to corpus content when generating queries, unlike real users. To mitigate this discrepancy, they restrict annotators to document summaries and metadata rather than the full source content. Text and image descriptions are first extracted from PDFs, summarized by an LLM, and then clustered through embedding and dimensionality reduction. Selected single-section and cross-section summaries are provided to annotators as the basis for query creation. ViDoRe V3 also supplements human-written queries with LLM-generated synthetic queries. Future benchmarks could further reduce source-induced bias by adopting query collection procedures that more closely reflect real-world information-seeking conditions.

\paratitle{Coverage of relevant items.} We next examine whether benchmarks attempt to identify all relevant pages or documents for each query. Under our criteria, only a small number explicitly assess relevance across the entire corpus and therefore support \emph{coverage of all relevant items}. For example, REAL-MM-RAG~\cite{REAL-MM-RAG} applies an MLLM to every page in the corpus to determine whether it contains sufficient information to answer a query. Queries are retained only when no page other than the annotated target is identified as answer-bearing, yielding a dataset in which each query is associated with a single verified relevant page. The benchmark further samples top-ranked retrieval results for human review to detect potential false-negative annotations.

More scalable approaches approximate corpus-wide assessment by first constructing a bounded candidate set through retrieval, clustering, knowledge-graph traversal, or model-based pre-screening, followed by human or model-assisted relevance annotation. Examples include DocViRe~\cite{DocViRe}, ViDoRe V3~\cite{ViDoRe-v3}, PaperScope~\cite{PaperScope}, and MultiDocR~\cite{MultiDocR}. MultiDocR, for instance, uses a retrieval-reranking-annotation pipeline: a retriever generates candidate pages, a reranker refines the set, and an MLLM assigns graded relevance labels. Such approaches improve scalability and reduce the likelihood of missing relevant items.

Beyond explicit corpus-wide relevance assessment, several benchmarks adopt indirect strategies to reduce annotation ambiguity and improve label reliability. A common approach is to make queries sufficiently discriminative so that the intended target is unlikely to be confused with other candidates. Some benchmarks achieve this by incorporating multiple entities or constraints into each query~\cite{Invoice-Haystack,CHaystack}, while others rely on LLMs to enforce specificity. For example, IRPAPERS~\cite{IRPAPERS} prompts the model to generate questions that are answerable by a unique page and to provide supporting justification, whereas SeaDoc~\cite{SeaDoc} filters generated queries based on their relevance and exclusivity to the designated target.

Other benchmarks apply post-hoc validation after query construction. IRPAPERS~\cite{IRPAPERS}, for instance, evaluates query discriminativeness by replacing the gold document with a hard negative and measuring the resulting change in downstream RAG performance. Although performance degradation suggests that the annotated document contains more useful evidence, it does not eliminate the possibility of false negatives, since the alternative document may still contain partially relevant information. KoViDoRe~\cite{KoViDoRe} instead uses a consistency-based filtering strategy, retaining a retrieved page as relevant only when it belongs to the same document from which the query was generated. While this improves annotation consistency, it may overlook relevant pages from other documents.

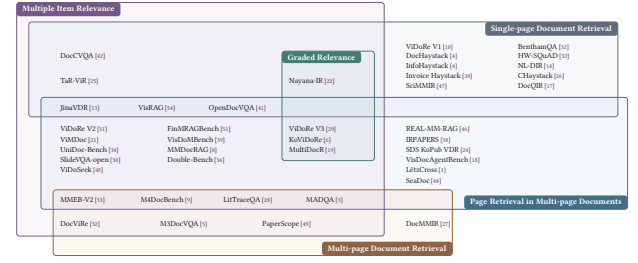
\begin{figure}[t]
\centering
\resizebox{0.98\columnwidth}{!}{\begingroup\rmfamily
\definecolor{pageink}{HTML}{436C88}
\definecolor{pagefill}{HTML}{91B7D2}
\definecolor{mpdocink}{HTML}{8D6346}
\definecolor{mpdocfill}{HTML}{E6B185}
\definecolor{multiink}{HTML}{735A86}
\definecolor{multifill}{HTML}{BCA5CC}
\definecolor{gradeink}{HTML}{477C70}
\definecolor{gradefill}{HTML}{8CBEB3}
\definecolor{spdocink}{HTML}{606C7E}
\definecolor{spdocfill}{HTML}{A8B5C7}

\begin{tikzpicture}[x=1bp,y=-1bp,
 setbox/.style={rounded corners=6bp,line width=1.8bp,fill opacity=.09},
 title/.style={rounded corners=6bp,text=white,inner xsep=12bp,inner ysep=6bp,font=\fontsize{17bp}{20bp}\selectfont\bfseries},
 item/.style={anchor=base west,inner sep=0pt,outer sep=0pt,text=black!90,font=\fontsize{16bp}{19bp}\selectfont}]
\path[use as bounding box] (-50,5) rectangle (1215,535);
\draw[setbox,draw=spdocink,fill=spdocfill] (-20,50) rectangle (1190,245);
\draw[setbox,draw=pageink,fill=pagefill] (5,205) rectangle (1210,435);
\draw[setbox,draw=mpdocink,fill=mpdocfill] (30,395) rectangle (850,530);
\draw[setbox,draw=multiink,fill=multifill] (-45,10) rectangle (710,490);
\draw[setbox,draw=gradeink,fill=gradefill] (500,110) rectangle (690,335);
\node[title,fill=multiink,anchor=north west] at (-45,10) {Multiple Item Relevance};
\node[title,fill=gradeink,anchor=north west] at (500,110) {Graded Relevance};
\node[title,fill=spdocink,anchor=north east] at (1190,50) {Single-page Document Retrieval};
\node[title,fill=pageink,anchor=south east] at (1210,435) {Page Retrieval in Multi-page Documents};
\node[title,fill=mpdocink,anchor=south east] at (850,530) {Multi-page Document Retrieval};
\node[item] at (45,470) {DocViRe\,{\fontsize{13bp}{15bp}\selectfont\cite{DocViRe}}};
\node[item] at (755,470) {DocMMIR\,{\fontsize{13bp}{15bp}\selectfont\cite{DocMMIR}}};
\node[item] at (755,105) {ViDoRe V1\,{\fontsize{13bp}{15bp}\selectfont\cite{ViDoRe-v1}}};
\node[item] at (45,275) {ViDoRe V2\,{\fontsize{13bp}{15bp}\selectfont\cite{ViDoRe-v2}}};
\node[item] at (515,275) {ViDoRe V3\,{\fontsize{13bp}{15bp}\selectfont\cite{ViDoRe-v3}}};
\node[item] at (755,275) {REAL-MM-RAG\,{\fontsize{13bp}{15bp}\selectfont\cite{REAL-MM-RAG}}};
\node[item] at (45,296) {ViMDoc\,{\fontsize{13bp}{15bp}\selectfont\cite{ViMDoc}}};
\node[item] at (755,296) {IRPAPERS\,{\fontsize{13bp}{15bp}\selectfont\cite{IRPAPERS}}};
\node[item] at (755,317) {SDS KoPub VDR\,{\fontsize{13bp}{15bp}\selectfont\cite{SDS-KoPub-VDR}}};
\node[item] at (755,338) {VisDocAgentBench\,{\fontsize{13bp}{15bp}\selectfont\cite{VisDocAgentBench}}};
\node[item] at (515,296) {KoViDoRe\,{\fontsize{13bp}{15bp}\selectfont\cite{KoViDoRe}}};
\node[item] at (45,317) {UniDoc-Bench\,{\fontsize{13bp}{15bp}\selectfont\cite{UniDoc-Bench}}};
\node[item] at (755,359) {L\"{e}tzCross\,{\fontsize{13bp}{15bp}\selectfont\cite{LetzCross}}};
\node[item] at (45,230) {JinaVDR\,{\fontsize{13bp}{15bp}\selectfont\cite{JinaVDR}}};
\node[item] at (755,125) {DocHaystack\,{\fontsize{13bp}{15bp}\selectfont\cite{DocHaystack-InfoHaystack}}};
\node[item] at (755,145) {InfoHaystack\,{\fontsize{13bp}{15bp}\selectfont\cite{DocHaystack-InfoHaystack}}};
\node[item] at (755,165) {Invoice Haystack\,{\fontsize{13bp}{15bp}\selectfont\cite{Invoice-Haystack}}};
\node[item] at (45,338) {SlideVQA-open\,{\fontsize{13bp}{15bp}\selectfont\cite{SlideVQA-DSE}}};
\node[item] at (205,230) {VisRAG\,{\fontsize{13bp}{15bp}\selectfont\cite{VisRAG}}};
\node[item] at (350,230) {OpenDocVQA\,{\fontsize{13bp}{15bp}\selectfont\cite{OpenDocVQA}}};
\node[item] at (250,470) {M3DocVQA\,{\fontsize{13bp}{15bp}\selectfont\cite{M3DocVQA}}};
\node[item] at (45,359) {ViDoSeek\,{\fontsize{13bp}{15bp}\selectfont\cite{ViDoSeek}}};
\node[item] at (265,275) {FinMRAGBench\,{\fontsize{13bp}{15bp}\selectfont\cite{FinMRAGBench}}};
\node[item] at (265,296) {VisDoMBench\,{\fontsize{13bp}{15bp}\selectfont\cite{VisDoMBench}}};
\node[item] at (755,185) {SciMMIR\,{\fontsize{13bp}{15bp}\selectfont\cite{SciMMIR}}};
\node[item] at (265,317) {MMDocRAG\,{\fontsize{13bp}{15bp}\selectfont\cite{MMDocRAG}}};
\node[item] at (985,105) {BenthamQA\,{\fontsize{13bp}{15bp}\selectfont\cite{BenthamQA-HW-SQuAD}}};
\node[item] at (985,125) {HW-SQuAD\,{\fontsize{13bp}{15bp}\selectfont\cite{BenthamQA-HW-SQuAD}}};
\node[item] at (45,125) {DocCVQA\,{\fontsize{13bp}{15bp}\selectfont\cite{DocCVQA}}};
\node[item] at (45,420) {MMEB-V2\,{\fontsize{13bp}{15bp}\selectfont\cite{MMEB-V2}}};
\node[item] at (985,145) {NL-DIR\,{\fontsize{13bp}{15bp}\selectfont\cite{NL-DIR}}};
\node[item] at (265,338) {Double-Bench\,{\fontsize{13bp}{15bp}\selectfont\cite{Double-Bench}}};
\node[item] at (515,317) {MultiDocR\,{\fontsize{13bp}{15bp}\selectfont\cite{MultiDocR}}};
\node[item] at (210,420) {M4DocBench\,{\fontsize{13bp}{15bp}\selectfont\cite{M4DocBench}}};
\node[item] at (460,470) {PaperScope\,{\fontsize{13bp}{15bp}\selectfont\cite{PaperScope}}};
\node[item] at (380,420) {LitTraceQA\,{\fontsize{13bp}{15bp}\selectfont\cite{LitTraceQA}}};
\node[item] at (755,380) {SeaDoc\,{\fontsize{13bp}{15bp}\selectfont\cite{SeaDoc}}};
\node[item] at (515,175) {Nayana-IR\,{\fontsize{13bp}{15bp}\selectfont\cite{Nayana-IR}}};
\node[item] at (985,165) {CHaystack\,{\fontsize{13bp}{15bp}\selectfont\cite{CHaystack}}};
\node[item] at (550,420) {MADQA\,{\fontsize{13bp}{15bp}\selectfont\cite{MADQA}}};
\node[item] at (985,185) {DocQIR\,{\fontsize{13bp}{15bp}\selectfont\cite{DocQIR}}};
\node[item] at (45,175) {TaR-ViR\,{\fontsize{13bp}{15bp}\selectfont\cite{TaR-ViR}}};
\end{tikzpicture}
\endgroup}
\caption{Categorization of VDR benchmarks by relevance annotation and ranking evaluation. Colored regions denote different retrieval settings, while overlaps indicate benchmarks supporting multiple settings.}
\label{fig:vdr-cardinality-ranking}
\Description{An overlapping-box diagram categorizing visual document retrieval benchmarks along five characteristics: multiple-item relevance, graded relevance, single-page document retrieval, page retrieval in multi-page documents, and multi-page document retrieval. Each labeled box contains benchmarks supporting the corresponding characteristic, while overlapping regions indicate benchmarks covering multiple characteristics. The diagram illustrates differences in retrieval granularity and relevance annotations across existing benchmarks, as well as their shared coverage.}
\end{figure}

\subsection{Cardinality of results and ranking}
\label{ssec:cardinality}

The third group of dimensions examines the number of annotated \emph{relevant items per query}, whether \emph{relevance level} is binary or graded, and the \emph{types of ranking supported}: page-level, document-level, or both. Together, these properties determine which aspects of our formulation a benchmark can evaluate.

To provide a clearer overview of how existing benchmarks support different retrieval settings, we present a Venn diagram in Figure~\ref{fig:vdr-cardinality-ranking}, organized by relevance annotation and ranking evaluation.

\paratitle{Number of relevant items per query.}
The number of \emph{relevant items per query} has evolved from predominantly single-pair supervision toward multi-positive and evidence-set annotations. Many benchmarks inherit a single answer-bearing page, figure, table, or image from existing QA or caption-pairing datasets~\cite{ViDoRe-v1,VisRAG,SciMMIR,NL-DIR,DocHaystack-InfoHaystack}, or generate queries from preselected target items~\cite{IRPAPERS,REAL-MM-RAG,SDS-KoPub-VDR,LetzCross,Invoice-Haystack}. Benchmarks annotating multiple relevant items can be broadly divided into two categories according to how relevance is defined and evaluated, see Appendix~\ref{sec:Rel_RAG_and_VDR}.

The first category follows the traditional IR setting, in which each query-item pair is annotated independently without explicitly modeling relationships among relevant items. Such annotations are most appropriate for the independent-item retrieval paradigm discussed in \S\ref{ssec:paradigms}. Representative benchmarks include DocViRe~\cite{DocViRe} and LitTraceQA~\cite{LitTraceQA} for document-level retrieval, and ViDoRe V2/V3~\cite{ViDoRe-v2,ViDoRe-v3} and MultiDocR~\cite{MultiDocR} for page-level retrieval.\footnote{As discussed in Appendix~\ref{sec:Rel_RAG_and_VDR}, some benchmarks incorporate multi-hop queries but retain item-wise relevance annotations and evaluation protocols. We therefore classify them under this category.} This setting aligns with conventional search scenarios, such as web search, where multiple independently relevant items can serve as alternative retrieval results.  

The second category focuses on complementary evidence retrieval. 
Benchmarks hence assess whether the retrieved results collectively provide sufficient evidence to answer the query. For example, Double-Bench~\cite{Double-Bench} defines multi-hop retrieval success as retrieving at least one supporting page for every reasoning hop within the top-$K$ results, measured by Hit@$K$. Although this setting imposes additional requirements on retrieval success, it still involves annotating multiple relevant items per query, typically spanning at least two distinct pages or documents. We therefore include it within the broader category of benchmarks supporting multiple relevant items per query. 

These categories reflect two distinct retrieval paradigms. However, both fit naturally within the retrieval framework introduced in \S~\ref{subsec:granularities}, which returns a ranked list of relevant items. Their primary distinction lies in the evaluation criterion: retrieved items are judged either independently by individual relevance or collectively according to the completeness of the evidence they provide.

\paratitle{Binary or graded relevance.} Most VDR benchmarks adopt binary judgments, while only few very recent benchmarks provide graded relevance annotations. This is partly attributable to the prevalence of single-positive annotations. Among benchmarks with graded relevance, ViDoRe V3~\cite{ViDoRe-v3} and KoViDoRe~\cite{KoViDoRe} distinguish among irrelevant, key-relevant, and sufficient-relevant pages. Nayana-IR~\cite{Nayana-IR} also adopts a three-level scheme, assigning the highest relevance to the exact target page, partial relevance to structurally related pages from the same parent document, and zero relevance otherwise. MultiDocR~\cite{MultiDocR} introduces a finer-grained five-level scheme, distinguishing fully, partially relevant, inference-relevant, topically relevant, and irrelevant pages. Collectively, these schemes move beyond binary relevance judgments to characterize how directly and sufficiently each item addresses a query. 

We advocate a flexible annotation setting that allows a variable number of relevant items per query, with graded judgments capturing their different degrees of relevance. Such a setting avoids assuming a fixed number of relevant results or treating all relevant items as equally sufficient to satisfy the query.

\begin{table*}[htbp]
  \centering
  \small
  \caption{Dimensions for assessing VDR benchmark readiness. The framework consolidates considerations commonly encountered during benchmark development and evaluation, providing a structured basis for comparing benchmark characteristics and design choices.}
  \label{tab:scoringCriteria}
  \begin{tabularx}{\textwidth}{@{} p{0.02\textwidth} X @{}}
    \toprule
    \rowcolor{criteriaBlue}
    \multicolumn{2}{@{}l}{\textbf{Sources and Structural Traceability (SS\&T)}} \\
    1 & The released data contain only isolated page images, images, figures, tables, page regions, or reassembled text blocks, and neither their source documents nor page order can be recovered. \\
    2 & The document to which each page belongs can be inferred, but document identifiers, page identifiers, or page order are incomplete. Pages have been extracted, removed, translated, re-rendered, or reorganized without adequate documentation. \\
    3 & Complete document boundaries are preserved, documents and pages have stable identifiers, and page order is explicit. The benchmark documents operations such as re-rendering, optical character recognition, translation, deduplication, filtering, and cropping. \\

    \rowcolor{criteriaBlue}
    \multicolumn{2}{@{}l}{\textbf{Corpus Size (CS): total number of pages across all documents}} \\
    1 & The underlying corpus contains fewer than 10,000 pages in total. \\
    2 & The underlying corpus contains at least 10,000 pages in total. \\

    \rowcolor{criteriaBlue}
    \multicolumn{2}{@{}l}{\textbf{Document Length (DL): number of pages per document}} \\
    1 & Corpus mainly consists of single-page documents. \\
    2 & The corpus consists mostly, if not entirely, of multi-page documents. \\

    \rowcolor{criteriaGreen}
    \multicolumn{2}{@{}l}{\textbf{Annotation Process (AP)}} \\
    1 & Before writing the query, the author sees a designated target page or a preselected fine-grained target evidence and constructs the query from it. \\
    2 & The query is constructed from one or more complete source documents, but the author is not directed to a particular target page or local region. The query is more independent than a question written from a designated page, yet it remains driven by candidate documents. \\
    3 & The query author can see only intentionally obscured corpus information, such as titles, metadata, abstracts, document summaries, descriptions of tables of contents, or the topic of the document collection, and cannot inspect specific evidence pages. Relevant items are searched for and annotated in the underlying document collection only after the query has been generated. \\
    4 & The query originates from real search logs, retrieval-system usage records, or other real information needs independently formulated by a human, and it predates relevance annotation. No target-corpus information is provided when the query is posed; relevance is annotated against the corpus only after the query has been issued. \\

    \rowcolor{criteriaGreen}
    \multicolumn{2}{@{}l}{\textbf{Coverage of Relevant Items (RC)}} \\
    1 & No attempt is made to ensure all potentially relevant items are labeled. \\
    2 & At least one of the following holds: (1) discriminative queries are designed to have a single gold candidate; or (2) checking is restricted to a limited subset, e.g., pages in the same document as the gold page. \\
    3 & A rule or model automatically examines every candidate without human verification.\textsuperscript{\dag} \\
    4 & Human verification is incorporated into the workflow. At least one of the following conditions holds: (1) one or more retrieval systems filter out low-relevance items and return top-ranked candidates for human judgment; (2) after a rule or model screens all candidates, humans verify a pre-specified random or stratified sample, estimate error rates, and determine whether the results are reliable; or (3) humans inspect every page of all documents in the corpus. \\

    \rowcolor{criteriaPurple}
    \multicolumn{2}{@{}l}{\textbf{Relevant Items per Query (RQ)}} \\
    1 & Each query has only single annotated relevant item. \\
    2 & The data allow both single-positive and multi-positive queries, but fewer than 20\% of queries have $\geq$2 relevant items. \\
    3 & At least 20\% but fewer than 50\% of queries have $\geq$2 directly annotated relevant items, and the full-dataset mean is $\leq$2 relevant items per query. \\
    4 & At least 50\% of queries have $\geq$2 directly annotated relevant items, or the full-dataset mean is $\geq$2 relevant items per query. \\

    \rowcolor{criteriaPurple}
    \multicolumn{2}{@{}l}{\textbf{Relevance Level (RL)}} \\
    1 & The data define binary relevant and non-relevant labels. \\
    2 & The annotation scheme recognizes partial relevance between full relevance and non-relevance, yielding three ordered relevance grades. \\
    3 & The data provide at least four ordered relevance grades, each with an operational definition and decision examples. \\

    \rowcolor{criteriaPurple}
    \multicolumn{2}{@{}l}{\textbf{Types of Ranking Supported (RS)}} \\
    1 & The benchmark evaluates page-level retrieval, including collections in which single-page content is treated as a document. Each page serves as an independent indexing unit, and no cross-page context modeling is required. Additional cases include document-level retrieval constructed directly from page-level annotations. \\
    2 & The benchmark supports document-level retrieval, in which queries are built upon the topical information of large portions of the document. It also covers contextual page-level retrieval, where accurate retrieval requires surrounding context rather than isolated pages. \\
    \bottomrule
  \end{tabularx}

  \medskip
  \begin{minipage}{\textwidth}
  \footnotesize
  \textsuperscript{\dag}Although a multimodal LLM with a carefully designed prompt can act as a human expert, human verification remains essential because humans are responsible for final data quality. Therefore, if the LLM outputs receive no human examination, the case falls into this level.
  
  \end{minipage}
  
\end{table*}

\paratitle{Page- or document-level ranking.}
Page-level retrieval remains dominant in VDR benchmarks. A common special case treats self-contained single-page images as nominal ``documents''~\cite{ViDoRe-v1,DocHaystack-InfoHaystack,Invoice-Haystack,SciMMIR,BenthamQA-HW-SQuAD,NL-DIR,CHaystack,Nayana-IR,TaR-ViR,DocQIR}. Although such settings technically support document-level retrieval, they do not evaluate selection among intact multi-page documents. Furthermore, most benchmarks based on multi-page documents treat pages as independent retrieval units rather than modeling cross-page dependencies. As a result, they offer limited assessment of document-level retrieval and reasoning over information distributed across multiple pages.

An emerging direction that addresses this limitation is \emph{contextual retrieval}, in which information from intermediate pages is required to identify the target page, even when those pages provide little direct evidence for answering the query. CMDR~\cite{CMDR} introduces this setting for intra-document retrieval, while VisDocAgentBench~\cite{VisDocAgentBench} extends it to VDR. In VisDocAgentBench, queries provide an initial semantic clue without directly specifying the target, requiring the retriever to navigate through supporting pages linked by latent semantic relationships. Although these pages may not contribute directly to answer generation, they act as essential intermediaries for locating the final evidence. The benchmark is explicitly designed for iterative retrieval, making it particularly suitable for agentic approaches that interleave planning, retrieval, and reasoning across multiple steps rather than relying on a single retrieval pass.

Contextual retrieval differs from conventional multi-hop evidence retrieval. In multi-hop retrieval, multiple pages jointly provide the evidence needed to answer a query. In contextual retrieval, by contrast, intermediate pages primarily help locate the target evidence. As noted in~\cite{CMDR}, many existing multi-hop queries explicitly mention concepts associated with individual evidence pages, allowing those pages to be retrieved directly from the query without exploiting dependencies between pages. Consequently, strong performance on this parallel multi-hop retrieval does not necessarily indicate contextual retrieval capability.

For document-level retrieval, native document-level supervision remains relatively uncommon, with some document-level evaluation protocols derived directly from page-level annotations. For example, MMEB-V2~\cite{MMEB-V2} infers document-level relevance from page-level labels, even though its queries target localized page content rather than document-level information. More direct document-level formulations include whole-document topical retrieval in DocViRe~\cite{DocViRe} and research-paper discovery in PaperScope~\cite{PaperScope}.

Beyond page- and document-level ranking, recent benchmarks have begun to incorporate finer-grained evidence localization. ViDoRe V3~\cite{ViDoRe-v3} augments page relevance with region grounding, M4DocBench~\cite{M4DocBench} links relevant documents and pages to supporting layouts, LitTraceQA~\cite{LitTraceQA} associates pages with evidence quotations, and BenthamQA and HW-SQuAD~\cite{BenthamQA-HW-SQuAD} combine page retrieval with answer-snippet localization.

Overall, existing VDR benchmarks remain dominated by independent page-level ranking. Native document-level supervision and fine-grained evidence localization are comparatively rare, while contextual retrieval highlights the need to evaluate dependencies across pages rather than only the direct relevance between individual pages and queries.

\section{Benchmark Readiness Scores and Findings}
\label{sec:scoring}

We now introduce a structured framework for characterizing and comparing VDR benchmarks. The framework is not intended to define a single notion of an "ideal" benchmark, and the resulting scores should not be interpreted as evidence of universal superiority across retrieval scenarios. Rather, its goal is to provide a systematic lens through which benchmark design choices can be examined. It serves two complementary purposes. First, it brings together considerations that are often discussed in isolation, including corpus construction, annotation methodology, relevance granularity, and application alignment. Second, it makes explicit the trade-offs inherent in benchmark development. By synthesizing practices observed across existing VDR benchmarks, the framework provides a common vocabulary for describing benchmark characteristics, identifying potential gaps or biases, and facilitating more transparent comparisons. Accordingly, the resulting scores should be viewed as a structured summary of benchmark properties rather than as an absolute ranking, recognizing that different research objectives and deployment settings may prioritize different dimensions.

Using this framework, we evaluate 42 benchmarks against the eight criteria defined in Table~\ref{tab:scoringCriteria}, which operationalize the dimensions introduced in \S\ref{sec:benchmarkReadiness}. Each criterion is scored on an ordinal scale comprising two to four levels. Initial assessments are generated by Claude Opus 5.5 and supported by evidence from the corresponding benchmark publication (e.g., direct quotations or section references) or from inspection of the benchmark annotation files. Two authors subsequently review all assessments independently and reach consensus on the final scores. The resulting evaluations are reported in Table~\ref{tab:benchmark-scores}, where benchmarks are ranked by their total score, computed as the unweighted sum of the eight criteria.

\begin{table}[t]
  \centering
  \small
  \caption{Recommended VDR benchmarks for six evaluation settings defined by retrieval granularity and paradigm. Scores (out of 24) are shown in parentheses.}
  \label{tab:rec-gran-para}
  \begin{tabularx}{\linewidth}{l|lL}
    \toprule
    \textbf{Granularity} & \textbf{Paradigm} & \textbf{Benchmarks} \\
    \midrule
    ~  & Items & ViDoRe V3 (21), MultiDocR (19), KoViDoRe (19), MMDocRAG (18), LitTraceQA (17), ViDoRe V2 (16)\\ \cmidrule{2-3}
    Page & Set & Double-Bench (20) \\ \cmidrule{2-3}
    ~  & Iterative & FinMRAGBench (16), VisDocAgentBench (16), M4DocBench (15), MADQA (15)\\
    \midrule
    ~ & Items & DocViRe (18), LitTraceQA (17) \\ \cmidrule{2-3}
    Document & Set & M3DocVQA (15)\\ \cmidrule{2-3}
    ~ & Iterative & PaperScope (18), M4DocBench (15)\\
    \bottomrule
  \end{tabularx}
\end{table}

\paratitle{Recommendations by granularity $\times$ retrieval paradigm.} As discussed in \S\ref{subsec:granularities} and \S\ref{ssec:paradigms}, VDR tasks vary along two dimensions: \emph{retrieval granularity} (page-level or document-level) and \emph{retrieval paradigm} (independent-item, evidence-set, or iterative retrieval). We organize existing benchmarks into six evaluation settings defined by the $2 \times 3$ combinations of these dimensions and provide benchmark recommendations for each setting in Table~\ref{tab:rec-gran-para}.

Existing benchmarks are predominantly concentrated in page-level independent-item retrieval. For this setting, we recommend several of the highest-scoring benchmarks: ViDoRe V3~\cite{ViDoRe-v3}, MultiDocR~\cite{MultiDocR}, KoViDoRe~\cite{KoViDoRe}, MMDocRAG~\cite{MMDocRAG}, LitTraceQA~\cite{LitTraceQA}, and ViDoRe V2~\cite{ViDoRe-v2}. These benchmarks generally feature more comprehensive relevance annotations, with several supporting multiple relevant items per query. Among them, ViDoRe V3, MultiDocR, and KoViDoRe further provide graded relevance judgments, enabling more nuanced evaluation of retrieval quality.

The remaining five settings are comparatively underrepresented, reflecting the relatively recent interest in document-level, evidence-set, and iterative retrieval within VDR research. We therefore include nearly all benchmarks identified for these settings. For page-level evidence-set retrieval, Double-Bench~\cite{Double-Bench} offers a well-designed evaluation protocol that requires retrieving complementary evidence across multiple reasoning hops. For page-level iterative retrieval, FinMRAGBench~\cite{FinMRAGBench}, VisDocAgentBench~\cite{VisDocAgentBench}, M4DocBench~\cite{M4DocBench}, and MADQA~\cite{MADQA} provide diverse evaluation settings that differ in their retrieval workflows and evidence-acquisition requirements.

Document-level retrieval benchmarks are even more limited. For independent-item retrieval, DocViRe~\cite{DocViRe} provides native document-level supervision through whole-document topical relevance. LitTraceQA derives document-level relevance from page-level evidence annotations. The latter also supports document-level evidence-set evaluation, alongside M3DocVQA~\cite{M3DocVQA}, whose document-level evidence sets can be derived from localized evidence annotations. In particular, M3DocVQA can be evaluated using set-level metrics such as All@$K$, as adopted in~\cite{HiKEY}. Finally, PaperScope~\cite{PaperScope} and M4DocBench provide evaluation environments for document-level iterative retrieval.

Overall, the recommended benchmarks cover all six evaluation settings, although their availability and annotation quality vary substantially. In particular, the scarcity of benchmarks beyond page-level independent-item retrieval highlights the need for more comprehensive evaluation resources supporting diverse retrieval granularities and paradigms. Beyond these two dimensions, we also summarize benchmarks featuring non-English queries or documents in Appendix~\ref{sec:non-english_bench}, providing additional guidance for researchers interested in evaluating retrieval performance across different languages.

\section{Conclusion}

This paper argues that visual document retrieval should be viewed as a family of retrieval tasks. We explicitly distinguishing retrieval granularity (page versus document) from retrieval paradigm, i.e., independent-item, evidence-set, and iterative retrieval. We assessed 42 existing benchmarks across eight dimensions covering corpus construction, annotation methodology, relevance coverage, and evaluation design. 

Our analysis reveals a clear pattern: benchmark development has focused primarily on independent page-level retrieval, while realistic query provenance, corpus-wide relevance assessment, graded relevance judgments, native whole-document supervision, evidence-set evaluation, and iterative retrieval remain comparatively rare. We believe the most important challenge for the next generation of VDR benchmarks is not increasing scale alone. Future benchmarks should move closer to authentic user information needs, explicitly characterize the modality and completeness of supporting evidence, and evaluate retrieval under settings where relevant information may be distributed across multiple pages, documents, and retrieval steps.


\bibliographystyle{ACM-Reference-Format}
\bibliography{references}


\appendix

\begin{table*}[t!]
  \centering
  \small
  \caption{Benchmark requirements for four representative applications, defined using the criteria in Table~\ref{tab:scoringCriteria}, and benchmarks satisfying these requirements based on the scores in Table~\ref{tab:benchmark-scores}.}
  \label{tab:app-requirements}
  \begin{tabularx}{\textwidth}{@{} >{\raggedright\arraybackslash}p{0.15\textwidth} L L L L @{}}
    \toprule
    \textbf{Application} & \textbf{Granularity} & \textbf{Paradigm} & \textbf{Selection rule} & \textbf{Benchmarks} \\
    \midrule
    Enterprise and knowledge-base search & Page; document for holistic requests & Independent items & SS=3, CS=2, DL=2, RC$\geq$3, RL$\geq$2 & ViDoRe V3, MultiDocR \\
    \midrule
    RAG over visually-rich documents & Page; Document & All three paradigms& DL=2, RC$\geq$3, RQ$\geq$2 & ViDoRe V3, Double-Bench, MultiDocR, MMDocRAG, DocViRe, PaperScope \\
    \midrule
    Financial and legal analysis & Page; Document & All three paradigms & SS=3, DL=2, RC$\geq$2, domain-specific & ViDoRe V3, MultiDocR, DocViRe, FinMRAGBench, M4DocBench \\
    \midrule
    Multi-document, multi-hop QA & Both, with cross-document aggregation & Evidence sets; iterative & DL=2, RC$\geq$3, RS=2 & Double-Bench, PaperScope, VisDocAgentBench \\
    \bottomrule
  \end{tabularx}
\end{table*}

\begin{table}[t]
  \centering
  \small
  \caption{Benchmarks with non-English queries/documents. Languages are represented by ISO 639-1 codes.}
  \label{tab:non-English_benchmarks}
  \begin{tabular}{l|ll}
  \toprule
    \textbf{Benchmarks} & \textbf{Query} & \textbf{Document} \\
    \midrule
    SDS KoPub VDR~\cite{SDS-KoPub-VDR} & ko & ko\\
    KoViDoRe~\cite{KoViDoRe} & ko & ko\\
    CHaystack~\cite{CHaystack} & zh & zh\\
    ViDoRe V1~\cite{ViDoRe-v1} & \emph{en}, fr & \emph{en}, fr\\
    DocQIR~\cite{DocQIR} & \emph{en}, hi, ja, ko, pt & \emph{en}\\
    JinaVDR~\cite{JinaVDR} & \emph{en} \& 19 langs.\textsuperscript{\dag} & \emph{en} \& 19 langs.\textsuperscript{\dag}\\
    L\"{e}tzCross~\cite{LetzCross} & \emph{en}, de, fr, lb & lb \\
    MMEB-V2~\cite{MMEB-V2} & \emph{en}, de, es, fr & \emph{en}, fr \\
    SeaDoc~\cite{SeaDoc} & \emph{en} & th, vi, ms, lo \\
    Nayana-IR~\cite{Nayana-IR} & \emph{en} \& 21 langs.\textsuperscript{\ddag} & \emph{en} \& 21 langs.\textsuperscript{\ddag} \\
    ViDoRe V2~\cite{ViDoRe-v2} & \emph{en}, de, es, fr & \emph{en} \\
    M4DocBench~\cite{M4DocBench} & \emph{en}, zh & \emph{en}, zh \\
    Double-Bench~\cite{Double-Bench} & \emph{en}, ar, es, fr, ja, zh & \emph{en}, ar, es, fr, ja, zh \\
    ViDoRe V3~\cite{ViDoRe-v3} & \emph{en}, de, es, fr, it, pt & \emph{en}, fr \\
    \bottomrule
  \end{tabular}

  \medskip
  
  \begin{minipage}{\textwidth}
  \footnotesize
  \textsuperscript{\dag}\emph{en}, ar, bn, de, es, fr, hi, hu, id, it, ja, ko, my, nl, pt, ru, th, ur, vi, zh

  \textsuperscript{\ddag}\emph{en}, ar, bn, de, es, fr, gu, hi, it, ja, kn, ko, ml, mr, or, pa, ru, sa, ta, te, th, zh
  
  \end{minipage}
\end{table}

\section{Applications and Benchmarks}
\label{sec:app-bench}

\subsection{Applications}
\label{ssec:applications}

We identify four application classes that VDR supports, and for each, justify why the underlying real-world need genuinely requires multi-page, visually-grounded retrieval, and which granularity(ies) it depends on. We also briefly justify why conventional text passage retrieval may not be the most suitable choice to support such applications. These justifications also parameterize the benchmark dimensions to be discussed shortly; Table~\ref{tab:app-requirements} summarizes the requirements each application places on a benchmark.

\paratitle{Enterprise and knowledge-base search.}
Organizations store institutional knowledge in PDFs, slide decks, and scanned reports whose relevant content is frequently non-textual: a pricing table, an architecture diagram, or a chart of quarterly results. Many such reports are long documents with tens of even hundreds of pages. A user searching an internal knowledge base needs the exact page containing the answer, not a document-level pointer that still requires manual skimming. This application is primarily \emph{page-level}: it requires large, heterogeneous document corpora with page-granular gold labels, and its practical value is proportional to how large and visually diverse the underlying corpus is. A complementary \emph{document-level} need arises for holistic requests such as ``find the deck that introduces product X'', the contract that is a lease, or the paper that surveys a topic (literature discovery, e-discovery). Here no single page need be relevant by itself, so page scores can only approximate document relevance (\S\ref{subsec:granularities}).

\paratitle{RAG over visually-rich documents.}
Retrieval-Augmented Generation (RAG) systems that answer questions over PDF corpora fail silently when their retriever discards visual evidence. A chart, a signature block, or a table cell could be the only place an answer lives, and no amount of generation quality compensates for a retriever that never surfaces it~\cite{VisRAG}. In practice this application needs \emph{both} granularities jointly. Page-level retrieval supplies the generator with the specific evidence pages. Document-level retrieval determines which whole documents are even worth pulling into the retrieval pool for answer generation, which matters once a corpus grows beyond what fits in a single context window~\cite{M3DocVQA}. Without document-level filtering, page-level retrieval alone may not come with the necessary context for citation-backed generation, critical for enterprise corpora.

\paratitle{Financial and legal document analysis.}
Financial filings, contracts, and regulatory reports are simultaneously long, numerically dense, and high-stakes, in the sense that an incorrect or ungrounded answer has direct legal or financial consequences. Practitioners in this domain routinely need two distinct capabilities: \emph{which} filings or contracts are relevant to a query, and \emph{where}, within a relevant document, the specific clause, number, or table appears. The former is document-level ranking, since an analyst may need to triage hundreds of candidate filings. The latter is page-level ranking, since citing the correct source location is often a regulatory or professional requirement. Text-only retrieval is a particularly poor fit here because financial and legal documents rely heavily on tabular and structured layout to convey numerical relationships.

\paratitle{Multi-document, multi-hop question answering.}
Some information needs cannot be answered from a single page or even a single document. An example use case is comparing a metric across several companies' annual reports, or reconciling a claim made in one document against supporting evidence in another. This application requires both granularities combined with cross-document aggregation: document-level ranking narrows a large corpus down to the relevant subset of documents, and page-level ranking then identifies the specific evidence within each. It is probably the most demanding application in terms of retrieval infrastructure.

\subsection{Recommendations Based on Applications}
\label{ssec:rec-app}

Table~\ref{tab:app-requirements} provides illustrative benchmark recommendations for different applications based on the scores in Table~\ref{tab:benchmark-scores}. The requirements are not prescriptive, and researchers may adjust the thresholds according to their specific priorities and evaluation needs.

\begin{table*}[t]
\centering
\caption{Characterization of 42 VDR benchmarks through the proposed readiness dimensions.  Scores should be interpreted as indicators of benchmark characteristics and design trade-offs rather than definitive rankings of benchmark quality.}
\label{tab:benchmark-scores}

\small
\setlength{\tabcolsep}{2.5pt}

\begin{tabularx}{\textwidth}{
    @{}
    >{\raggedright\arraybackslash}p{0.15\textwidth}
    *{10}{>{\centering\arraybackslash}X}
    @{}
}

\toprule
\textbf{Dimensions}
& \multicolumn{3}{c}{\textbf{Document source and characteristics}}
& \multicolumn{2}{c}{\textbf{Query and Annotation}}
& \multicolumn{3}{c}{\textbf{Results and Ranking}}
& \textbf{Total}  & \textbf{Retrieval} \\
\cmidrule(lr){2-4}
\cmidrule(lr){5-6}
\cmidrule(lr){7-9}

\textbf{Benchmark}
&
\cellcolor{criteriaBlue}\textbf{SS\&T}
&
\cellcolor{criteriaBlue}\textbf{CS}
&
\cellcolor{criteriaBlue}\textbf{DL}
&
\cellcolor{criteriaGreen}\textbf{AP}
&
\cellcolor{criteriaGreen}\textbf{RC}
&
\cellcolor{criteriaPurple}\textbf{RQ}
&
\cellcolor{criteriaPurple}\textbf{RL}
&
\cellcolor{criteriaPurple}\textbf{RS}
&
\textbf{Score}
&
\textbf{Paradigm}
\\

\midrule

ViDoRe V3~\cite{ViDoRe-v3}
& 3 & 2 & 2 & 3 & 4 & 4 & 2 & 1 & 21 & Items\\

Double-Bench~\cite{Double-Bench}
& 3 & 2 & 2 & 2 & 4 & 4 & 1 & 2 & 20 & Set\\

MultiDocR~\cite{MultiDocR}
& 3 & 2 & 2 & 1 & 3 & 4 & 3 & 1 & 19 & Items\\

KoViDoRe~\cite{KoViDoRe}
& 3 & 1 & 2 & 3 & 3 & 4 & 2 & 1 & 19 & Items\\

MMDocRAG~\cite{MMDocRAG}
& 3 & 2 & 2 & 1 & 4 & 4 & 1 & 1 & 18 & Items\\ 

DocViRe~\cite{DocViRe}
& 3 & 2 & 2 & 2 & 4 & 2\textsuperscript{\dag} & 1 & 2 & 18 & Items\\

PaperScope~\cite{PaperScope}
& 3 & 1\textsuperscript{\dag} & 2 & 2 & 3 & 4 & 1 & 2 & 18 & Iterative\\

LitTraceQA~\cite{LitTraceQA}
& 3 & 2 & 2 & 2 & 2 & 4 & 1 & 1 & 17 & Items\\

FinMRAGBench~\cite{FinMRAGBench}
& 3 & 2 & 2 & 1 & 2 & 4 & 1 & 1 & 16 & Iterative\\

VisDocAgentBench~\cite{VisDocAgentBench}
& 3 & 1 & 2 & 2 & 4 & 1 & 1 & 2 & 16 & Iterative\\

ViDoRe V2~\cite{ViDoRe-v2}
& 3 & 1 & 2 & 3 & 1 & 4 & 1 & 1 & 16 & Items\\

M4DocBench~\cite{M4DocBench}
& 3 & 1 & 2 & 1 & 2 & 4 & 1 & 1 & 15 & Iterative\\

M3DocVQA~\cite{M3DocVQA}
& 3 & 2 & 2 & 1 & 1 & 4 & 1 & 1 & 15 & Set\\

MADQA~\cite{MADQA}
& 3 & 2 & 2 & 2 & 2 & 2 & 1 & 1 & 15 & Iterative\\

UniDoc-Bench~\cite{UniDoc-Bench}
& 3 & 2 & 2 & 1 & 1 & 4 & 1 & 1 & 15 & Items \\

ViDoSeek~\cite{ViDoSeek}
& 3 & 1 & 2 & 1 & 3 & 3 & 1 & 1 & 15 & Items\\

SDS KoPub VDR~\cite{SDS-KoPub-VDR}
& 3 & 2 & 2 & 1 & 3 & 1 & 1 & 1 & 14 & Items \\

REAL-MM-RAG~\cite{REAL-MM-RAG}
& 3 & 1 & 2 & 1 & 4 & 1 & 1 & 1 & 14 & Items \\

DocCVQA~\cite{DocCVQA}
& 3 & 1 & 1 & 1 & 2 & 4 & 1 & 1 & 14 & Items \\ 

TaR-ViR~\cite{TaR-ViR}
& 1 & 2 & 1 & 4 & 1 & 2 & 1 & 1 & 13 & Items \\

Nayana-IR~\cite{Nayana-IR}
& 2 & 2 & 1 & 1 & 2 & 2 & 2 & 1 & 13 & Items \\

VisDoMBench~\cite{VisDoMBench}
& 3 & 2 & 2 & 1 & 2 & 1 & 1 & 1 & 13 & Items \\

ViMDoc~\cite{ViMDoc}
& 3 & 2 & 2 & 1 & 1 & 2 & 1 & 1 & 13 & Items\\

CHaystack~\cite{CHaystack}
& 3 & 2 & 1 & 1 & 2 & 1 & 1 & 1 & 12 & Items\\

SeaDoc~\cite{SeaDoc}
& 3 & 1 & 2 & 1 & 2 & 1 & 1 & 1 & 12 & Items\\

MMEB-V2~\cite{MMEB-V2}
& 2 & 2 & 1 & 1 & 2 & 2 & 1 & 1 & 12 & Items\\

SlideVQA-open~\cite{SlideVQA-DSE}
& 2 & 2 & 2 & 1 & 1 & 2 & 1 & 1 & 12 & Items\\

L\"{e}tzCross~\cite{LetzCross}
& 3 & 1 & 2 & 1 & 2 & 1 & 1 & 1 & 12 & Items\\

IRPAPERS~\cite{IRPAPERS}
& 3 & 1 & 2 & 1 & 2 & 1 & 1 & 1 & 12 & Items\\

Invoice Haystack~\cite{Invoice-Haystack}
& 3 & 1 & 1 & 1 & 2 & 1 & 1 & 1 & 11 & Items\\

InfoHaystack~\cite{DocHaystack-InfoHaystack}
& 1 & 1 & 1 & 1 & 4 & 1 & 1 & 1 & 11 & Items\\

DocHaystack~\cite{DocHaystack-InfoHaystack}
& 1 & 1 & 1 & 1 & 4 & 1 & 1 & 1 & 11 & Items\\

DocMMIR~\cite{DocMMIR}
& 1 & 2 & 1 & 2 & 1 & 1 & 1 & 2 & 11 & Items\\

OpenDocVQA~\cite{OpenDocVQA}
& 1 & 2 & 1 & 1 & 1 & 2 & 1 & 1 & 10 & Items\\

VisRAG~\cite{VisRAG}
& 1 & 2 & 1 & 1 & 1 & 2 & 1 & 1 & 10 & Items\\

BenthamQA~\cite{BenthamQA-HW-SQuAD}
& 2 & 1 & 1 & 1 & 1 & 1 & 1 & 1 & 9 & Items\\

SciMMIR~\cite{SciMMIR}
& 1 & 2 & 1 & 1 & 1 & 1 & 1 & 1 & 9 & Items\\

JinaVDR~\cite{JinaVDR}
& 1 & 2 & 1 & 1 & 1 & 1 & 1 & 1 & 9 & Items\\

DocQIR~\cite{DocQIR}
& 1 & 1 & 1 & 1 & 1 & 1 & 1 & 1 & 8 & Items\\

NL-DIR~\cite{NL-DIR}
& 1 & 1 & 1 & 1 & 1 & 1 & 1 & 1 & 8 & Items\\

HW-SQuAD~\cite{BenthamQA-HW-SQuAD}
& 1 & 1 & 1 & 1 & 1 & 1 & 1 & 1 & 8 & Items\\

ViDoRe V1~\cite{ViDoRe-v1}
& 1 & 1 & 1 & 1 & 1 & 1 & 1 & 1 & 8 & Items\\

\bottomrule

\end{tabularx}

\medskip
\begin{minipage}{\textwidth}
\footnotesize
\textsuperscript{\dag}When information is insufficient for precise scoring, the dataset receives the lowest score among cases consistent with its paper's description.

\end{minipage}

\end{table*}

\section{Relationship between VDR and RAG} \label{sec:Rel_RAG_and_VDR}

This section explains how the evolving requirements of RAG have shaped the three VDR retrieval paradigms introduced in \S\ref{ssec:paradigms}.

RAG enhances large language models (LLMs) by incorporating external information into the answer generation process. In a typical RAG pipeline, a retriever identifies relevant content from a knowledge corpus, and a generator uses the retrieved evidence to formulate a response. Because the quality of the retrieved content directly influences the relevance, completeness, and factual grounding of generated answers, retrieval is a critical component of RAG. From a page-centric indexing and search perspective, RAG can be viewed as a two-stage process comprising visual document retrieval to identify relevant pages, followed by answer generation.

Indeed, with the development of RAG, VDR has increasingly extended beyond retrieving individually relevant pages toward acquiring multiple pieces of evidence to support complex question answering. However, such multi-page retrieval can involve different objectives, leading to distinct retrieval and evaluation settings. Two key distinctions emerge: \textit{(i) whether relevance is assessed at the individual-item or evidence-set level}, and \textit{(ii) whether retrieval is conducted in a single pass or through an iterative process}. Based on these distinctions, we discuss three retrieval paradigms commonly considered in existing VDR benchmarks.

\paratitle{Conventional information retrieval (IR)}, where each retrieved item is evaluated independently according to its relevance to the query. Even when multiple pages or documents are relevant to the same query, their relevance labels are assigned individually, and retrieval is evaluated based on the resulting ranked list. Standard IR metrics are therefore applicable. For example, nDCG@$K$ is commonly used when graded relevance judgments are available, whereas Recall@$K$, MAP, and MRR are widely adopted when relevance is treated as binary.

\paratitle{Complementary evidence across multiple pages.} The second paradigm is motivated by RAG scenarios where answering a complex query requires combining complementary evidence distributed across multiple pages. The objective is to recover a sufficient set of evidence that \textit{collectively} supports answer generation. This requirement has motivated the introduction of multi-hop queries and corresponding set-level evaluation protocols. For example, M3DocVQA~\cite{M3DocVQA} provides query-level evidence-set annotations, allowing evaluation through metrics such as All@$K$, as adopted in \cite{HiKEY}, where retrieval succeeds only when the complete annotated evidence set is covered by the top-$K$ results. Notably, this metric was not part of M3DocVQA's original evaluation protocol. Double-Bench~\cite{Double-Bench} further accounts for the substitutability of evidence by defining multi-hop retrieval success as retrieving at least one supporting page for every reasoning hop within the top-$K$ results, measured by Hit@$K$. Nevertheless, the presence of multi-hop queries does not necessarily imply set-level retrieval evaluation. Several benchmarks, including KoViDoRe~\cite{KoViDoRe}, UniDoc-Bench~\cite{UniDoc-Bench}, and ViDoRe V3~\cite{ViDoRe-v3}, incorporate multi-hop questions but continue to adopt conventional item-wise IR protocols, although their definitions of multi-hop queries differ.

\paratitle{Iterative retrieval.} The recent emergence of agentic RAG has introduced a third paradigm: iterative retrieval. Unlike conventional RAG pipelines that typically retrieve evidence once before answer generation, agentic systems can interleave planning, retrieval, evidence inspection, and reasoning across multiple steps. Accordingly, evidence acquisition becomes a sequential and history-dependent process, where observations from earlier steps inform subsequent retrieval decisions and determine when sufficient evidence has been collected. Several VDR benchmarks explore this paradigm from different perspectives. VisDocAgentBench~\cite{VisDocAgentBench} introduces latent semantic bridges that must be resolved to locate the final target; MADQA~\cite{MADQA} emphasizes acquiring minimal grounded evidence from heterogeneous collections; M4DocBench~\cite{M4DocBench} incorporates conversational dependencies and annotated evidence chains; FinMRAGBench~\cite{FinMRAGBench} focuses on iterative financial analysis; and PaperScope~\cite{PaperScope} provides a tool-based environment for scientific information seeking. Additionally, these benchmarks do not share a unified evaluation protocol. For instance, VisDocAgentBench evaluates final-target retrieval using Recall and MRR, MADQA measures the quality of retrieved citation sets using F1, and PaperScope separately reports direct-retrieval Recall for Topic Induction and downstream reasoning accuracy.

Overall, RAG has broadened the objectives and evaluation settings of VDR, from ranking independently relevant items to recovering complementary evidence sets and supporting iterative evidence acquisition. These distinctions highlight the importance of selecting benchmarks that align with the intended retrieval objective, considering not only the document domains and corpus characteristics but also whether relevance is defined at the item or evidence-set level, whether relevant items are interchangeable or complementary, and whether evidence is acquired through single-shot or iterative retrieval.

\section{Benchmarks with Non-English Settings}\label{sec:non-english_bench}

Table~\ref{tab:non-English_benchmarks} summarizes VDR benchmarks featuring non-English queries or documents. Some benchmarks focus on individual non-English languages without multilingual or cross-lingual evaluation~\cite{SDS-KoPub-VDR,KoViDoRe,CHaystack}. The remaining benchmarks support diverse language settings, including translated queries~\cite{DocQIR,ViDoRe-v2,ViDoRe-v3,LetzCross}, multilingual document collections~\cite{SeaDoc}, and language-specific subsets~\cite{Nayana-IR,JinaVDR}. Researchers can select benchmarks according to their target languages and evaluation settings.


\end{document}